\documentclass[trackchanges,twocolumn]{aastex701}

\usepackage{graphicx}
\usepackage{lipsum}
\usepackage{subcaption}
\usepackage{placeins}
\usepackage{natbib}
\usepackage{hyperref}
\usepackage{xcolor}

\definecolor{red}{rgb}{1,0,0}

\begin{document}

\title{Correlation between baryonic process and galaxy assembly bias}

\correspondingauthor{Zhongxu Zhai}
\email{zhongxuzhai@sjtu.edu.cn}

\author{Zilan Xiao}
\affiliation{State Key Laboratory of Dark Matter Physics, Tsung-Dao Lee Institute \& School of Physics and Astronomy, Shanghai Jiao Tong University, Shanghai 200240, China}
\affiliation{Shanghai Key Laboratory for Particle Physics and Cosmology, and Key Laboratory for Particle Physics, Astrophysics and Cosmology, Ministry of Education, Shanghai Jiao Tong University, Shanghai 200240, China}
\email{zilanxiao@sjtu.edu.cn}

\author{Junyu Hua}
\affiliation{State Key Laboratory of Dark Matter Physics, Tsung-Dao Lee Institute \& School of Physics and Astronomy, Shanghai Jiao Tong University, Shanghai 200240, China}
\affiliation{Shanghai Key Laboratory for Particle Physics and Cosmology, and Key Laboratory for Particle Physics, Astrophysics and Cosmology, Ministry of Education, Shanghai Jiao Tong University, Shanghai 200240, China}
\email{junyu.hua@sjtu.edu.cn}

\author{Zhongxu Zhai}
\affiliation{State Key Laboratory of Dark Matter Physics, Tsung-Dao Lee Institute \& School of Physics and Astronomy, Shanghai Jiao Tong University, Shanghai 200240, China}
\affiliation{Shanghai Key Laboratory for Particle Physics and Cosmology, and Key Laboratory for Particle Physics, Astrophysics and Cosmology, Ministry of Education, Shanghai Jiao Tong University, Shanghai 200240, China}
\affiliation{Waterloo Center for Astrophysics, University of Waterloo, Waterloo, ON N2L 3G1, Canada}
\affiliation{Department of Physics and Astronomy, University of Waterloo, Waterloo, ON N2L 3G1, Canada}
\email{zhongxuzhai@sjtu.edu.cn}

\author{Andrew Benson}
\affiliation{Carnegie Observatories, 813 Santa Barbara Street, Pasadena, CA 91101, USA}
\email{abenson@carnegiescience.edu}

\author{Yun Wang}
\affiliation{IPAC, California Institute of Technology, Mail Code 314-6, 1200 E. California Blvd., Pasadena, CA 91125, USA}
\email{ywang@ipac.caltech.edu}

\received{Month DD, YYYY}
\revised{Month DD, YYYY}
\accepted{Month DD, YYYY}
\submitjournal{ApJ}

\begin{abstract}
Galaxy assembly bias (GAB) is the dependence of galaxy clustering on secondary properties beyond halo mass. In this work, we study the connections between GAB and baryonic processes using the Galacticus semianalytic model (SAM) for galaxy formation and evolution applied to the UNIT simulation. By generating hundreds of galaxy mocks with varying parameters governing gas cooling, star formation, stellar feedback, and AGN feedback, we quantify the GAB signal using a shuffling technique that compares the clustering of the original galaxy samples with that of shuffled samples, in which galaxies are randomly reassigned among haloes of the same mass. We then compare the GAB signals associated with halo concentration and smoothed environmental density, and use the Random Forest (RF) algorithm to identify which baryonic processes are most correlated with GAB within the explored SAM parameter space. We find that for stellar-mass-selected galaxies, the dominant baryonic processes are gas cooling and stellar feedback, and the result does not change significantly with the number density; for star formation rate (SFR)-selected galaxies, the primary process shifts from star formation to gas cooling as the number density increases. These results reveal a correlational dependence of GAB on baryonic processes within the explored Galacticus SAM parameter space.
\end{abstract}


\section{Introduction}
In the current paradigm of galaxy formation, galaxies form and evolve within dark matter halos, and their properties are intimately connected to their host halos. Modeling the connection between galaxies and halos not only helps to understand the physical processes of galaxy formation, but also helps to constrain cosmological parameters through galaxy surveys, such as SDSS-I/II \citep{York2000,Abazajian2009}, 2dFGRS \citep{Colless2001,Cole2005}, WiggleZ \citep{Drinkwater2010}, BOSS \citep{Dawson2013}, eBOSS \citep{Dawson2016}, DESI \citep{DESI2016,DESI2024}, Euclid \citep{Laureijs2011,Euclid2022}, and the Roman Space Telescope \citep{Spergel2015,Wang2022Roman}. In recent years, advances in large-scale galaxy surveys and high-resolution numerical simulations have led to the emergence of numerous models to describe the statistical galaxy--halo connection \citep{Wechsler2018}. In general, these models can be divided into two categories: empirical models, such as subhalo abundance matching (SHAM) \citep{Kravtsov2004,Tasitsiomi2004,Vale2004} and the halo occupation distribution (HOD) \citep{Jing1998,Benson2000,Peacock2000,Seljak2000,White2001,Berlind2002,Cooray2002,Zheng2005}; and physical models, such as hydrodynamic simulations \citep{Vogelsberger2014,Pillepich2018,Pakmor2023} and semianalytical models (SAM) \citep{White1991,Cole2000,Benson2010,Guo2013}. Due to the high-resolution and large-scale computing requirements, hydrodynamic simulations consume substantial computing resources, thereby limiting both the simulation volume and the ability to thoroughly explore the parameter space, although they can model galaxies in great detail. On the other hand, SAMs can approximate many astrophysical processes of galaxy formation and evolution with empirical prescriptions, thereby accelerating computations and overcoming the above limitations of fully exploring the parameter space.

In the excursion set formalism and the basic structure formation scenario, the clustering of dark matter halos is influenced solely by their mass \citep{Press1974,Bond1991}. However, it is observed in N-body simulations that halo clustering can also depend on other halo properties beyond mass, including formation time, concentration, and spin \citep{Gao2005,Wechsler2006,Gao2007,Li2008}. This phenomenon is now generally termed halo assembly bias (HAB). These additional properties are collectively referred to as secondary halo properties, and the dependency of clustering on them is referred to as secondary bias \citep{Mao2017}. Throughout this paper, we refer to this effect as assembly bias for simplicity.

In empirical models, such as the standard HOD, the occupation of galaxies in halos depends only on halo mass. However, the assembly history of halos not only affects the clustering of halos but also may affect the occupation of galaxies in halos, showing occupation variation (OV) (\citealt{Zehavi2018,Artale2018}). It denotes the differences in the occupation of galaxies residing in halos of the same mass. For example, older halos tend to have a higher probability of hosting a central galaxy at low masses; however, among halos of the same mass, older halos contain fewer satellite galaxies than younger ones \citep{Contreras2019}. This OV combined with HAB indicates that the clustering properties of galaxies are determined by host halo mass and secondary halo properties, i.e. galaxy assembly bias (GAB).

This phenomenon was noticed by \citet{Croton2007} using a SAM applied to the Millennium Simulation. \citet{Contreras2019} studied how the influence of the halo concentration and halo age on HAB, OV, and GAB evolves with redshift. The results show that the halo age-based GAB mainly comes from OV, while the concentration-based GAB mainly comes from HAB. \citet{Xu2021a} further analyzed the contributions of various secondary properties to GAB, including internal properties such as halo age, concentration, and angular momentum, as well as various external environmental factors. The results seem to show that the environmental density is the most important secondary property for GAB \citep{Hadzhiyska2020,Xu2021a,Yuan2021b}.

On the other hand, observational evidence for GAB remains inconclusive. Some studies have reported positive detections \citep{Cooper2010,Wang2013,Miyatake2016,Zentner2019,Obuljen2020,Yuan2021a,Yuan2021b}, while others have found the impact of assembly bias to be small or consistent with zero \citep{Blanton2007,Lin2016,Zu2016,Walsh2019,Salcedo2022,Shao_2025}. While the observational status of GAB remains debated, a number of studies have focused on parameterizing GAB in theoretical models to enable controllable mock catalogues and facilitate future detection. Previous studies have successfully incorporated GAB into empirical models, including composite abundance matching \citep{Lehmann2017,Contreras2021} and modified HOD models \citep{Hearin2016b,McEwen2018,Wibking2018,Yang2026}. These theoretical models of GAB rely on secondary halo properties as phenomenological proxies, while the connection between baryonic processes and the GAB remains largely unexplored. \citet{Daalen2014} found that AGN feedback removes material from the inner regions of subhaloes, lowering their central density concentration and making them less compact at fixed mass. Such internal structural changes affect the spatial distribution and clustering of subhaloes on small to intermediate scales (\citealt{Romeo_2020, Romeo_2023}). \citet{Yang2026} compared the redshift evolution of GAB in two hydrodynamical simulations, SIMBA and IllustrisTNG. They found that the divergent GAB trends at $z<2$ can be attributed to differences in AGN feedback implementation, suggesting that baryonic physics plays a crucial role in shaping GAB. This indicates that the impact of baryonic effects on GAB deserves further investigation.

As mentioned earlier, hydrodynamic simulations can be ideal to study the baryonic effect. However, it is not feasible to sample the parameter space with sufficient volume and obtain a robust clustering measurement. Therefore, we turn to SAMs and examine the impacts of various baryonic effects on the GAB signatures in this work. In SAM, galaxy formation and evolution are driven by processes including gas cooling, star formation, supernova feedback, active galactic nucleus (AGN) feedback, etc \citep{Benson2010,Wechsler2018}. Using the Galacticus SAM and a machine learning algorithm, we quantify the importance of individual baryonic processes for the GAB signal in galaxy samples selected with different secondary properties and number densities.

This paper is organized as follows. In Section \ref{sec: SIMULATION AND METHOD}, we introduce our data from the UNIT simulation and Galacticus SAM, as well as the secondary halo properties we used to isolate the effect of assembly bias. Section \ref{sec: IMPACT OF SECONDARY PROPERTIES ON GAB} presents our results of OV and GAB, showing their dependence on different secondary properties and galaxy number density, and the dependence of GAB on baryonic processes. Finally, we conclude with a summary and discussion in Section \ref{sec: Conclusions}.

\section{Simulation and method}
\label{sec: SIMULATION AND METHOD}
\subsection{UNIT simulation}
We build the SAM models using the UNIT cosmological simulation \citep{Chuang2019}. This simulation adopts the following cosmological parameters: \(\Omega_{\rm m}=0.3089\), \(h=0.6774\), \(n_{\rm s}=0.9667\), \(\sigma_8=0.8147\) \citep{Planck2015}, and evolves \(4096^{3}\) dark matter particles within a periodic cubic volume of side length \(1\,h^{-1}\mathrm{Gpc}\). The particle mass is \(1.2\times10^9\,h^{-1}M_\odot\), which can resolve halos down to \(10^{11}\,h^{-1}M_\odot\) corresponding to roughly 100 particles.  Dark matter halos are identified using the ROCKSTAR halo finder \citep{Behroozi2012a}, and merger trees are constructed with the Consistent Tree software \citep{Behroozi2012b}. With high mass resolution, large volume, and complete merger trees, these outputs enable mock catalogue construction using SAM and subsequent analysis of GAB.

\subsection{Galacticus Semianalytic Model}
We employ the Galacticus SAM \citep{Benson2012} to build our galaxy catalogues. Following \citet{Zhai2025}, we focus on the galaxy sample at \(z=0.55\), which is selected to match the BOSS CMASS compilation, but we note that previous studies have found that galaxy assembly bias evolves with redshift \citep
[e.g.,][]{Contreras2019}. The methodology applied in this work can be extended to other redshifts and galaxy selections. For simplicity, we use the products from our earlier work \citet{Zhai2025} and refer the reader to that work for more details on the SAM modeling; here we only provide a brief summary. In the Galacticus modeling of galaxy formation including star formation, gas cooling, AGN feedback, etc., we allow 16 parameters of these components to vary (see Table~2 in \citealt{Zhai2025}) and investigate their influence on the GAB signal. These parameter ranges provide a physically motivated sampling of the Galacticus model space, with the Latin-hypercube design in \citet{Zhai2025} excluding parameter combinations that produced unphysically high star formation. The parameter ranges span physically plausible values while allowing significant variation in the resulting galaxy properties. By sampling this multi-dimensional parameter space, we can systematically explore the influence of different baryonic processes on the galaxy-halo connection and the resulting GAB signal. The original suite contains 262 Galacticus mocks. After applying the HOD resolution test described in Appendix~\ref{sec:appendix_hod_resolution}, we use the remaining 133 mocks throughout the analyses.

For each SAM mock, we construct galaxy samples using two selection schemes. In the stellar-mass-selected sample, galaxies are rank-ordered by their stellar mass and the most massive galaxies are selected until the target number density is reached. In the SFR-selected sample, galaxies are rank-ordered by their star formation rate, and the galaxies with the highest SFR are selected in the same way. Note that due to the variation of the SAM predictions, the galaxy mocks can have different cuts on stellar mass or star formation rate with the same required number density.

\subsection{Secondary halo properties}
In the study of secondary bias of dark matter halos and galaxies, it is crucial to specify the relevant secondary halo properties. These include halo properties beyond mass, such as formation time, concentration, spin, and external environment of halos, which can impact the clustering of both galaxies and dark matter halos. In this study, we focus on two such properties: halo concentration and smoothed dark matter density to represent internal and external properties, respectively. We note that this choice is not exclusive; a study that fully explores a broader set of properties could provide a more comprehensive evaluation of assembly bias, as in \citet{Xu2021a}.

(1) Halo concentration \(c\) is defined as the ratio between the virial radius \(r_{\text{vir}}\) and the characteristic scale radius \(r_{\text{s}}\) of the Navarro--Frenk--White (NFW) density profile \citep{NFW1996} of a dark matter halo: \(c = r_{\text{vir}}/r_{\text{s}}.\)
The virial radius \(r_{\text{vir}}\) marks the boundary within which the halo is approximately virialized. The scale radius \(r_{\text{s}}\) marks the radial scale where the logarithmic slope of the NFW profile changes. A higher concentration indicates that the halo mass is more centrally concentrated.

(2) The smoothed environmental matter density \(\delta_R\) is measured using a top-hat window function of radius \(R\). We consider two smoothing scales, \(R = 10\,h^{-1}\mathrm{Mpc}\) and \(R = 5\,h^{-1}\mathrm{Mpc}\), denoted as \(\delta_{10}\) and \(\delta_5\), respectively. These quantities provide a simple but effective measure of the density-field environment around halos \citep{McEwen2018}.

\subsection{Method}
\label{sec: METHODOLOGY}
We quantify the clustering signals of the galaxies using the two-point correlation function \(\xi(r)\), which measures the excess probability of finding a galaxy pair with separation \(r\) compared with a random distribution. In order to avoid the complexities of redshift space distortion due to peculiar velocity, all the computation in our work is done in real space.

With the clustering measurements, a straightforward way of evaluating the GAB signal is to split the galaxy sample by halo properties of interest and compare the resultant clustering signal. Alternatively, we employ the shuffling method as widely used in literature (e.g., \citealt{Croton2007,Zu2008,Chaves2016,Contreras2019,Xu2021a}), i.e. compare the correlation function of the original galaxy sample with that of a shuffled sample. In particular, we shuffle the galaxies among halos within narrow bins of halo mass. The central galaxies are moved to the center of the new halos, and satellite galaxies are moved together with their central galaxies to keep their positions relative to the centrals. This procedure is able to retain the one-halo clustering term while removing correlations due to halo properties other than mass. The GAB signal is then defined as the difference in the two-point correlation functions before and after this shuffling procedure. In our analysis, the halo mass range of \(\log_{10} M_h \,[h^{-1} M_\odot] = 9.8\) to \(15.2\) is divided into equal-width bins of \(0.2\)~dex.

We also employ a \textit{double shuffle} method to isolate individual contributions of different secondary properties to the GAB signal (\citealt{Xu2021a}). In this case, we shuffle the galaxies within the same host halo mass and an additional secondary property (concentration, or environmental density \( \delta_{10} \), \( \delta_{5} \) ). The resulting GAB signal thus quantifies the specific influence of that particular secondary property. We use the same halo-mass bins as in the single shuffling, and further divide halos into bins of the chosen secondary property before shuffling. Concentration is also logarithmically binned, using 15 bins with a width of roughly \(0.2\). For \(\delta_5\), the bin widths are \(0.5\), \(1\) and \(2\) for the ranges \(0\)--\(6\), \(6\)--\(12\) and \(12\)--\(20\), respectively; for \(\delta_{10}\), the bin widths are \(0.2\) and \(0.5\) for the ranges \(0\)--\(6\) and \(6\)--\(11\), respectively. 

\section{Results}
\label{sec: IMPACT OF SECONDARY PROPERTIES ON GAB}
In this section, we present our main results. We first examine the halo occupation distribution (HOD) of the galaxy mock from our SAM. We then present our evaluation of GAB measurements. Finally, we adopt a random forest method to quantify the impact of various baryonic processes on the significance of GAB signals.

\subsection{Occupation variation}

\begin{figure*}
    \centering
    \includegraphics[width=1.0\textwidth]{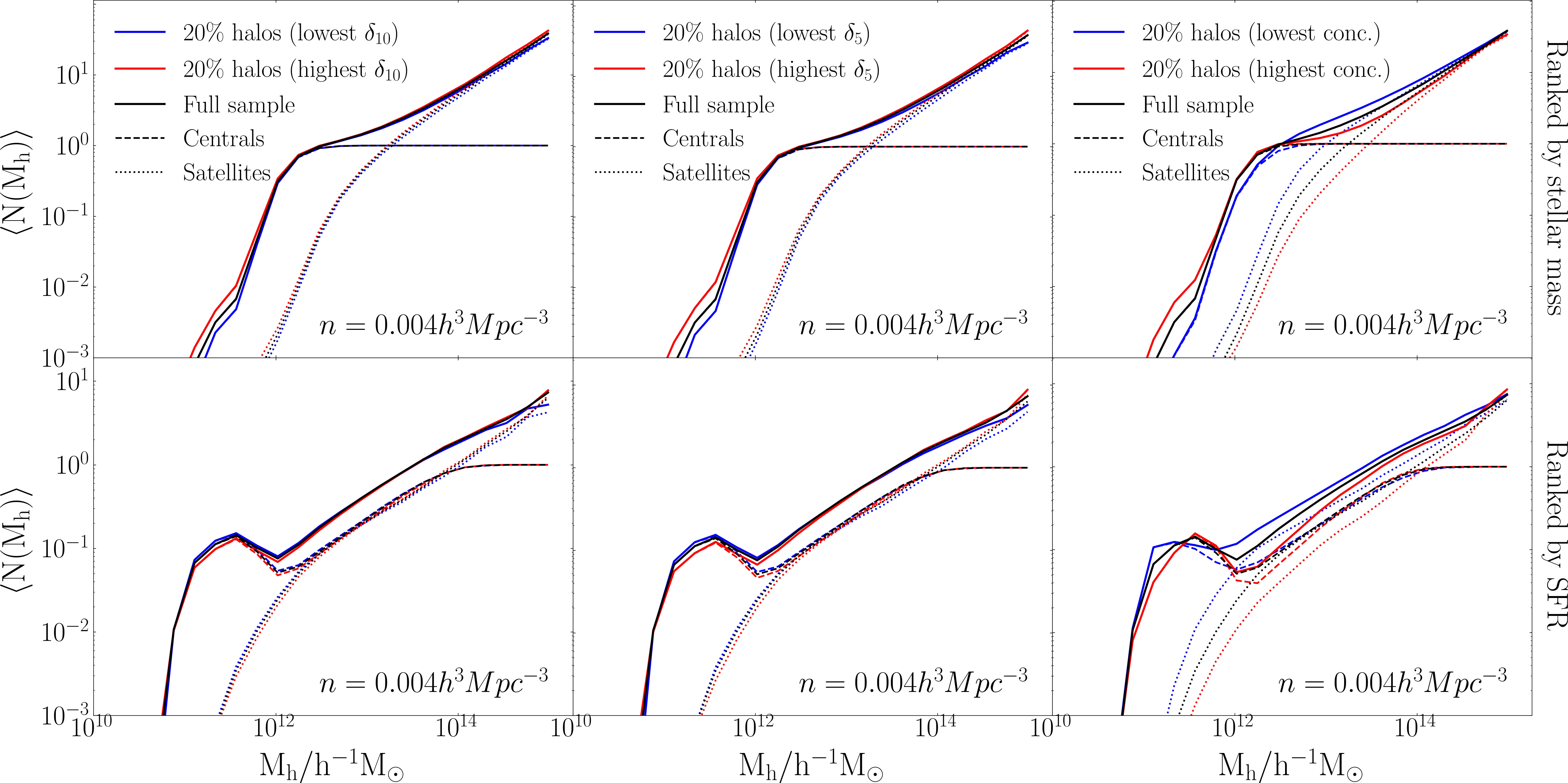} 
    \caption{Halo occupation distributions (HODs) for a galaxy mock with number density of \(n=0.004h^{3}\mathrm{Mpc}^{-3}\). The top and bottom panels correspond to galaxies selected by stellar mass and star formation rate (SFR), respectively. From left to right, the columns show results for \(\delta_{10}\), \(\delta_{5}\), and concentration, respectively. In each panel, red and blue lines correspond to halos with the highest and lowest 20 percent of the given property, while line styles (dashed, dotted, solid) distinguish central, satellites, and the total galaxies.}
    \label{fig:OV}
\end{figure*}

Since OV, together with HAB, constitutes the GAB, we first examine the OV to gain insight into the subsequent GAB measurements. To examine the OV with concentration and environmental density, we first rank halos within narrow bins of fixed mass according to a given secondary property, and then compare the HOD for the subsets of halos with the highest and lowest \(20\%\) of that property.

Fig.~\ref{fig:OV} illustrates how the halo occupation distribution is influenced by halo concentration and environmental density (\(\delta_5\), \(\delta_{10}\)). We generate hundreds of galaxy mocks by varying the physical parameters within the Galacticus SAM framework. The figure presents the results from only a single representative mock for illustration. The red (blue) curves correspond to the \(20\%\) of halos with the highest (lowest) concentration or environmental density. Dashed, dotted and solid curves denote the centrals, satellites and full sample.

For stellar-mass-selected centrals (top), the HOD of central galaxy rises monotonically from 0 to 1 with halo mass as expected, i.e. the most massive galaxies tend to live in the most massive dark matter halos \citep{White2011}. For SFR-selected centrals (bottom), the HOD exhibits a non-monotonic trend, i.e. star forming galaxies preferentially live in halos of intermediate masses, a phenomena that has been observed in multiple SAM mocks \citep{Zhai2019,Merson2019}.

The two panels in the rightmost column display the OV for halo concentration. The impact of concentration differs significantly between stellar-mass-selected and SFR-selected samples for central galaxies. For stellar-mass-selected centrals, high concentration enhances occupation at low masses but suppresses it at high masses, while for SFR-selected centrals, it suppresses occupation across the entire mass range. For satellite galaxies, the dependence on concentration is similar for both stellar-mass-selected and SFR-selected samples. High-concentration halos host fewer satellites than their low-concentration counterparts at fixed halo mass. These results are largely consistent with \citet{Contreras2019}. On average, the impact of concentration on OV is more pronounced for satellites than centrals, although this does not necessarily translate into a larger satellite contribution to GAB.

The four panels in the left and middle columns show the OV attributed to environmental density. For stellar-mass-selected samples, halos in denser environments begin to host central galaxies at slightly lower masses and contain slightly more satellites at fixed halo mass. The OV become negligible as the halo mass increases, which is consistent with \citet{Zehavi2018} and \citet{Xu2021a}. Compared to environmental density, concentration has a stronger impact on OV. This may be understood in the context of how secondary properties introduce scatter into the stellar mass--halo mass (SMHM) relation. \citet{Zehavi2018} found that formation time introduces more scatter to the SMHM relation for central galaxies than does environmental density, and such scatter is found to diminish at higher masses. The stronger OV we find for concentration may similarly reflect that concentration introduces greater scatter than environment. However, it is worth noting that the relative strength of OV does not directly or necessarily dictate the contribution to GAB. As will be shown in Sec.~\ref{sec: GAB}, despite its weaker OV, the environmental density contributes a stronger GAB signal than does concentration.

\subsection{Galaxy assembly bias}
\label{sec: GAB}

\begin{figure*}
    \includegraphics[width=1.0\textwidth]{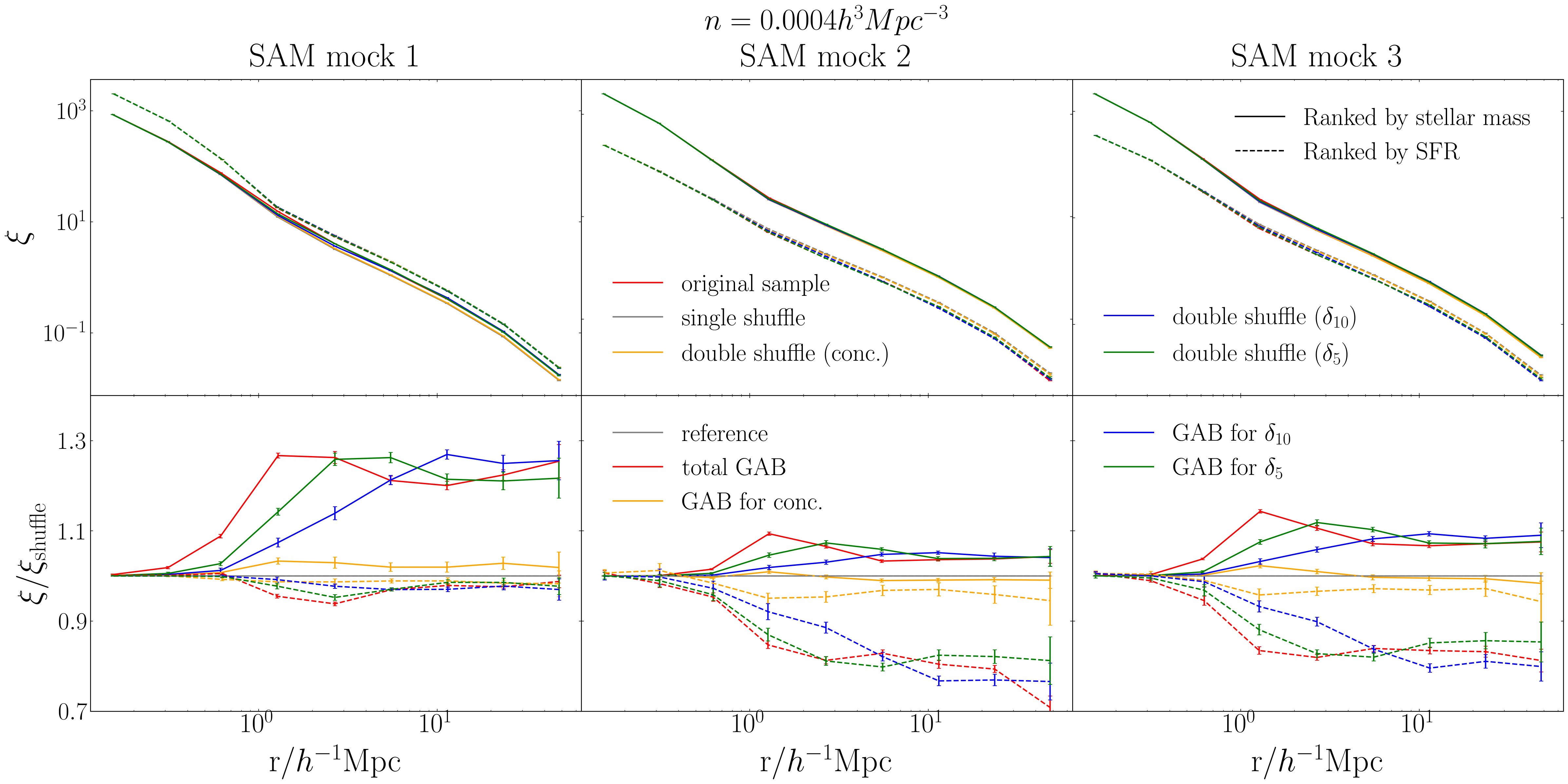} 
    \caption{The GAB signals attributed to different secondary properties for three SAM mocks (\(n = 0.0004h^{3}Mpc^{-3}\)). Top panel: two-point correlation functions. The colored lines represent the original sample and various shuffled samples. The dashed and solid lines correspond to stellar mass- and SFR-selected samples, respectively. Bottom panel: the ratio of correlation functions. The gray line represents the ratio of the correlation function with the single shuffle sample as the numerator, and its value is always 1 as a reference. The red line corresponds to the original sample, which retains total GAB signal. The yellow, blue, and green lines correspond to GAB associated with the halo concentration, environmental density \(\delta_{10}\) and \(\delta_{5}\). The error bars in the figure are derived from the standard deviation obtained by performing 10 random shuffle operations on the same galaxy sample, reflecting the statistical uncertainty introduced by the shuffling procedure.}
    \label{fig:ratio}
\end{figure*}

\begin{figure*}[htbp]
    \centering
    \includegraphics[width=1.0\textwidth]{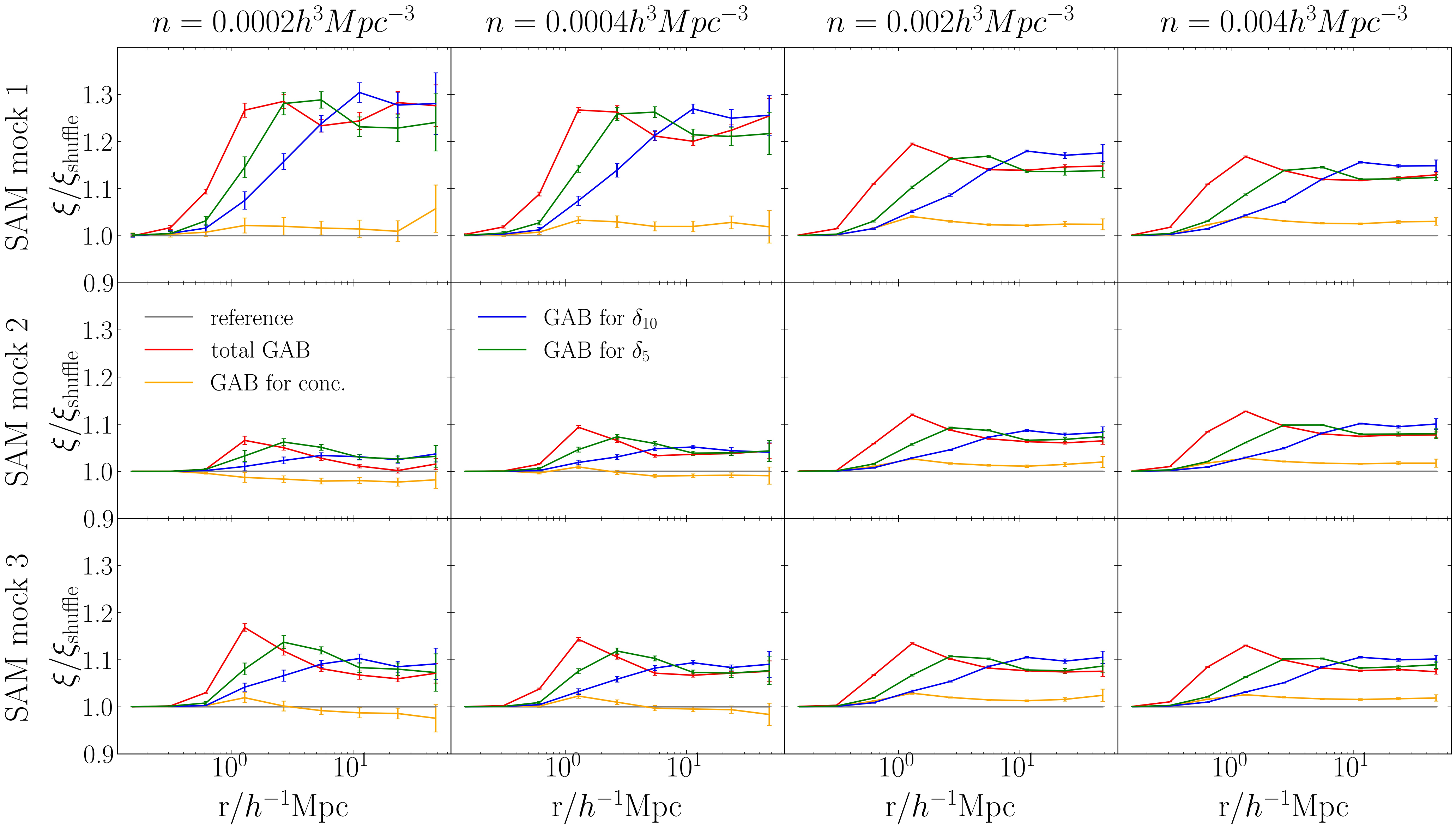} 
    \caption{GAB signals for stellar-mass-selected galaxy samples with different number densities. The figure shows the ratio of two-point correlation functions, following the same definition as in the bottom panels of Fig.~\ref{fig:ratio}. From left to right, the columns correspond to \(n = 0.0002h^{3}Mpc^{-3}\), \(n = 0.0004h^{3}Mpc^{-3}\), \(n = 0.002h^{3}Mpc^{-3}\), and \(n = 0.004h^{3}Mpc^{-3}\). The rows show SAM mocks 1, 2, and 3 from top to bottom. The colored curves represent the total GAB signal and the contributions associated with halo concentration, \(\delta_{10}\), and \(\delta_{5}\), as indicated in Fig.~\ref{fig:ratio}.}
    \label{fig:ratio_density_Mstar}
\end{figure*}

\begin{figure*}[htbp]
    \centering
    \includegraphics[width=1.0\textwidth]{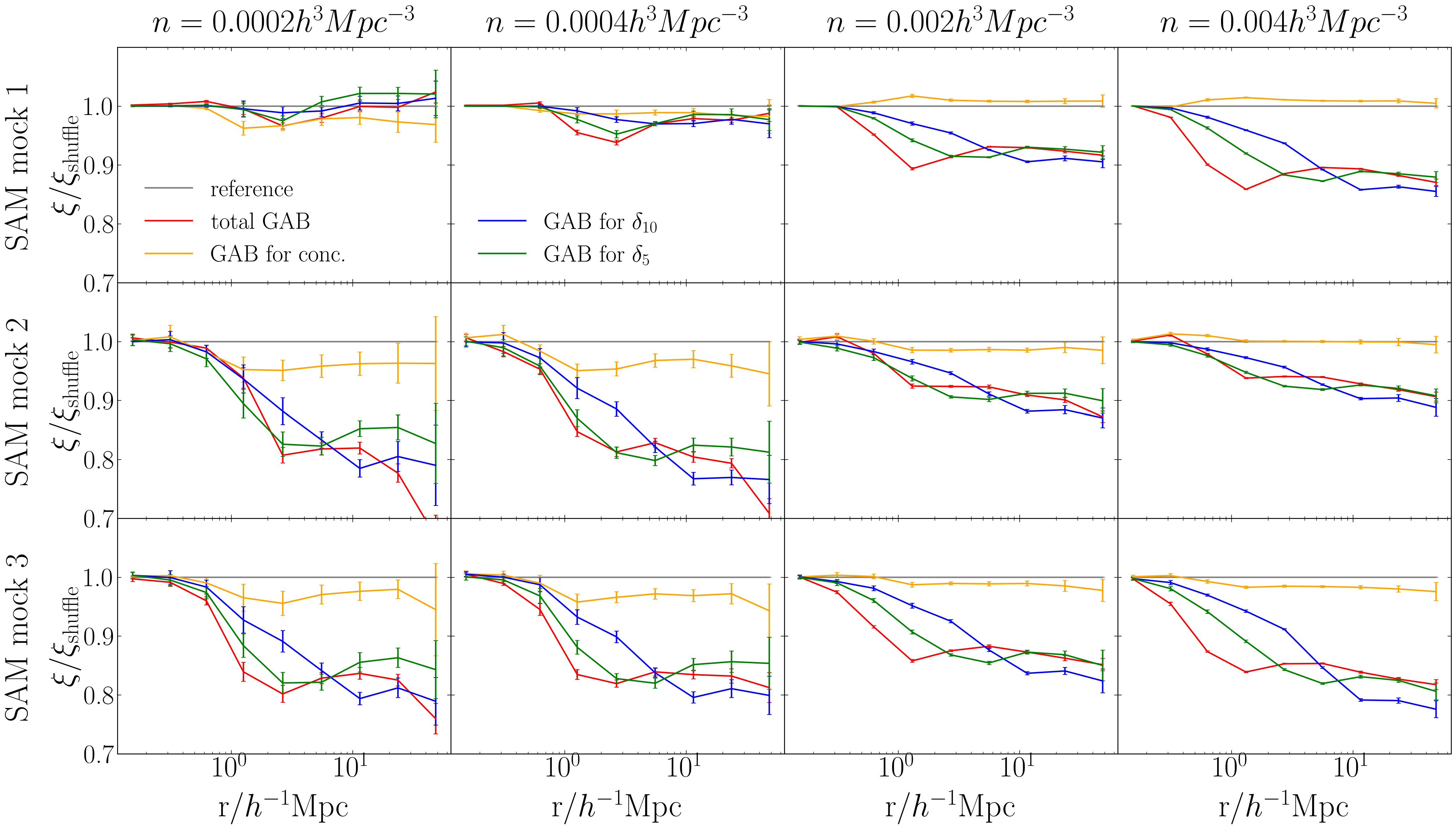} 
    \caption{Same as Fig.~\ref{fig:ratio_density_Mstar}, but for SFR-selected galaxy samples. }
    \label{fig:ratio_density_SFR}
\end{figure*}

We investigate the GAB signals with different sample selections. The single-shuffled sample, constructed by randomizing galaxy positions among halos of the same mass, serves as a reference devoid of GAB, as it breaks the correlation between galaxy positions and any secondary halo properties beyond mass. The total GAB signal is then quantified by the ratio of the two-point correlation functions from the original and the single-shuffled samples. Similarly, double-shuffled samples are used to isolate the GAB contribution from a specific secondary property (concentration or environmental density in this analysis). After the shuffling procedure, the HOD of the full sample (black lines in Fig. \ref{fig:OV}) remains unchanged, while the HODs of the subsamples ranked by concentration or environmental density (red and blue lines in Fig. \ref{fig:OV}) become identical to that of the full sample \citep{Contreras2019}. To estimate the statistical uncertainty of the shuffling procedure, we repeat the shuffling process 10 times for each galaxy sample and compute the standard deviation of the resulting GAB signals. These uncertainties are shown as error bars in the figures, reflecting the variation among the shuffle realizations, and we note that they may not represent the full measurement uncertainty of the clustering ratio.

Fig. \ref{fig:ratio} shows the two-point correlation functions and the GAB signals generated by different secondary properties. For illustration purposes, we present results from three representative mocks from our SAM runs. As shown later in Figs.~\ref{fig:ratio_density_Mstar} and \ref{fig:ratio_density_SFR}, these three mocks illustrate the main types of number-density dependence found in our SAM sample. The top panel shows the two-point correlation functions for the galaxy sample with number density \(n = 0.004h^{3}Mpc^{-3}\). The bottom panel plots the ratio of the two-point correlation functions, \(\xi / \xi_{\text{single shuffle}}\). Here, the denominator \(\xi_{\text{single shuffle}}\) represents a reference result with GAB signal erased. Different curves represent different choices for the numerator \(\xi\): the original sample (red) gives the total GAB; the single-shuffled sample itself (grey) yields a constant ratio of unity as a reference; and the double-shuffled samples (yellow, blue, green) isolate the GAB contributions from concentration, \(\delta_{10}\), and \(\delta_{5}\), respectively. The dashed and solid lines correspond to the result of stellar mass- and SFR-selected samples.

On small scales dominated by one-halo terms, the difference between the original and shuffled samples is negligible as expected \citep{Croton2007}, since the relative positions of galaxies within the same halo remain unchanged after shuffling. On large scales, we first note a difference between the selection criteria. For stellar-mass-selected samples, the clustering ratio is greater than unity, while for SFR-selected samples, it is less than unity \citep{Contreras2019}, indicating that GAB can in some cases reduce the clustering amplitude.

The representative mocks shown in Fig.~\ref{fig:ratio} also demonstrate that different SAM parameter choices can produce different clustering ratios, indicating that variations in baryonic physics can change the amplitude of the GAB signal. This behavior is similar to the findings in  \citet{Contreras2024}. Using hydrodynamical simulations with different physical implementations, they found that the amplitude of the GAB signal depends on the physical models of galaxy formation.

Regarding the contributions from different secondary properties, we find that for most of our SAM mocks, the GAB produced by \(\delta_{10}\) and \(\delta_{5}\) is close to the total GAB, while concentration only produces a small fraction, as exemplified by Fig.~\ref{fig:ratio}. This is consistent with earlier works such as \citet{Xu2021a} and \citet{Hadzhiyska2020}. We note that the distinction between intrinsic halo properties and external environment is important for interpreting these results. At fixed halo mass, one can in principle construct a galaxy sample without any GAB driven by concentration. However, it is unlikely to construct a galaxy sample without any dependence on environment in the same way, because quantities like $\delta_{5}$ or $\delta_{10}$ are direct measurements of the surrounding density field and are therefore intrinsically connected to clustering. This difference explains why the concentration-related assembly bias is not necessarily correlated with the environmental measurement: the two cases do not probe the same physical effect.

Figs.~\ref{fig:ratio_density_Mstar} and \ref{fig:ratio_density_SFR} illustrate the number-density dependence of the GAB signals for stellar-mass-selected and SFR-selected galaxy samples, respectively. Based on our examination of the full set of SAM mocks, these three models represent the main types of number density dependence observed in the parameter space, but we note that this is dependent on the prescriptions of the model and the ranges of the parameter space adopted in our Galacticus modeling.

Type I (mock 1) shows a decreasing clustering ratio with increasing number density for both stellar-mass- and SFR-selected samples, consistent with \citet{Xu2021a}, who reported a similar trend for stellar-mass-selected galaxies. Type II (mock 2) exhibits the opposite trend: the clustering ratio increases with number density for both selection criteria, with the SFR-selected sample even crossing from below unity to above unity. This pattern is similar to the findings in \citet{Contreras2019}. 
Type III (mock 3) shows weak number-density dependence, with clustering ratios remaining broadly stable across the four number-density thresholds. This diversity indicates that the response of GAB to number density varies across the SAM parameter space, ranging from decreasing to increasing trends. In addition, we note that the uncertainty increases at lower number densities due to the reduced sample size.

\subsection{Relative importance of Baryonic Physics on GAB}
\label{sec: Relative importance of Baryonic Physics on GAB}

As the previous figures show, different SAM mocks with the same number density and selection condition can predict GAB signals with different significance. This can be attributed to the strength of the astrophysical processes of each model. In this section, we explore the impact of these baryonic prescriptions on the resultant GAB signals.

Before we explore the contribution of each component, we must first define the strength of GAB signal. Following the early work by \cite{Wang2025}, we pick the ratio of the correlation function with and without shuffling at $r\sim10h^{-1}$Mpc as a proxy of the GAB measure. Given the high-dimensional space of our SAM parameters, there are multiple ways to evaluate the contribution of each parameter or physical process. For instance, we can define a fiducial model in the parameter space and then change one parameter each time and rerun the SAM. The comparison of the resultant mock and the fiducial one can provide a direct estimate of the difference. Alternatively, our analysis is based on the early work from \cite{Zhai2025} that has provided hundreds of SAM mocks with a Latin-hypercube sampling in the parameter space. In order to utilize these products without rerunning the SAM algorithm, we choose a different analysis strategy in this work.

We use a Random Forest (RF) regression algorithm \citep{Breiman2001} implemented in \texttt{scikit-learn} with permutation importance to analyze the impact of various galaxy formation parameters on the GAB signal. Random Forest is an ensemble method that builds multiple decision trees and aggregates their predictions. Beyond its predictive capabilities, it provides estimates of feature importance, which can measure how much each input variable contributes to explaining the variance in the target quantity.

Feature importance can be assessed through different approaches. The impurity-based method (also known as Mean Decrease in Impurity, MDI) \citep{Louppe2015} calculates the total reduction in node impurity attributable to each feature across all trees, which is the default method in \texttt{scikit-learn}. The permutation importance method \citep{Breiman2001} evaluates the importance of a feature by randomly shuffling the values of that feature and measuring the resulting drop in model performance. Unlike impurity-based importance, permutation importance does not overestimate the importance of features with many unique values, providing a more reliable estimate of each feature's contribution to the model's predictions. We therefore adopt permutation importance.

\begin{figure*}
    \centering
    \includegraphics[width=1.0\textwidth]{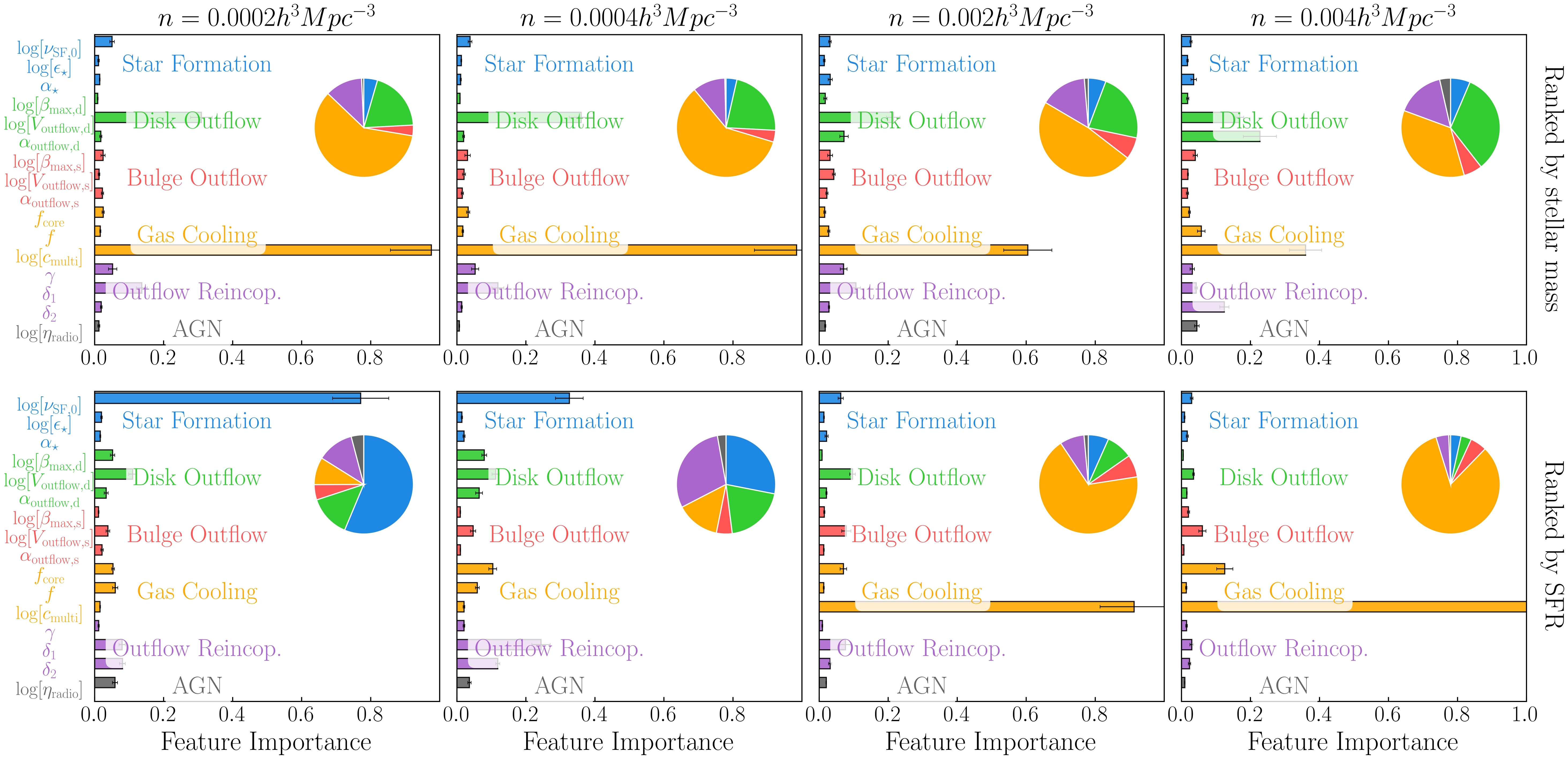} 
    \caption{The relative importance of various baryonic processes on total GAB. The top and bottom rows correspond to stellar-mass-selected and SFR-selected galaxy samples, respectively. From left to right, the columns show increasing galaxy number density. Physical processes, including star formation, stellar feedback, gas cooling, outflow reincorporation, and AGN feedback, are color-coded as shown in the legend. The error bars represent the standard deviation across 10 random permutations. The inset pie charts in each panel display the relative contribution of each physical process group to the total GAB}.
    \label{fig:feature_importance}
\end{figure*}

In our analysis, the input features are the 16 SAM parameters governing various baryonic processes, and the target variable is the amplitude of the GAB signal measured at a characteristic scale of \(\sim 10\,h^{-1}\) Mpc. These SAM parameters are grouped according to the baryonic processes they control: \(\log[\nu_{\rm SF,0}]\), \(\log[\epsilon_\star]\), and \(\alpha_\star\) for star formation; \(\log[\beta_{\max,d}]\), \(\log[V_{\rm outflow,d}]\), and \(\alpha_{\rm outflow,d}\) for stellar feedback in the disk; \(\log[\beta_{\max,s}]\), \(\log[V_{\rm outflow,s}]\), and \(\alpha_{\rm outflow,s}\) for  stellar feedback in the spheroid; \(f_{\rm core}\), \(f\), and \(\log[c_{\rm multi}]\) for gas cooling; \(\gamma\), \(\delta_1\), and \(\delta_2\) for outflow reincorporation; and \(\log[\eta_{\rm radio}]\) for AGN feedback. These process groups are color-coded in Fig.~\ref{fig:feature_importance}. We refer readers to Table~2 and Section 3.2 of \citet{Zhai2025} for the detailed definition of each individual SAM parameter. For each feature, we compute permutation importance with 10 random shuffles; the error bars in the resultant figures of RF analysis represent the standard deviation across these 10 repetitions, reflecting the uncertainty due to the permutation procedure. This approach allows us to rank the SAM parameters according to their impact on GAB. By comparing these importance rankings across galaxy samples selected by stellar mass and SFR, we can identify which baryonic processes play a more significant role in producing GAB than the others.

In Fig. \ref{fig:feature_importance}, we present the feature importance on total GAB. Additional feature importance results are shown in Appendix \ref{sec:appendix_rf}. The panels are arranged from left to right in order of increasing galaxy number density, revealing the difference between stellar-mass-selected (top) and SFR-selected (bottom) galaxies. Bar colors indicate different physical processes. The error bars represent the standard deviation across 10 random permutations, reflecting the stochastic uncertainty of the permutation importance procedure. The inset pie charts summarize the relative contribution of each group of baryonic processes to the total GAB signal.

For the stellar mass-selected samples, gas cooling (orange) and stellar feedback within the disk (green) play a dominant role in determining the strength of GAB. In addition, we find that this behavior does not change significantly as we increase the number density. Given the tight correlation between stellar mass and halo mass, the sample becomes more complete at the high-mass end. Therefore higher number density means the GAB signal is more dominated by less massive halos. Our analysis seems to show that baryonic effects contribute similarly across different halo masses in terms of GAB.

On the other hand, we find that for the SFR-selected samples, the contributions of baryonic effects can change significantly as we increase the number density. At low number density, the dominant process is gas cooling (orange) followed by star formation (blue). At high number density, the gas cooling process becomes the single dominant factor and the star formation parameters are no longer affecting the GAB signal significantly.

For the main RF feature-importance analysis shown in Fig.~\ref{fig:feature_importance}, we use all the SAM mocks for training. We also perform a five-fold cross-validation test to assess the predictive performance of the RF model. In this test, the mocks are divided into five subsets; in each fold, four subsets are used for training, and the remaining subset is used for validation. We find that the cross-validated \(R^2\) values are lower than those obtained when evaluating the model on the full training set, indicating limited predictive generalization and some degree of overfitting. However, the relative feature-importance rankings remain broadly consistent with those obtained using the full set of the mocks.

In addition to the RF analysis, we perform supplementary checks using Spearman correlation, Pearson correlation, and SHAP analyses (\citealt{SHAP_ref}). The results are broadly consistent with the RF analysis, confirming the robustness of our findings. In Appendix \ref{sec:appendix_Spearman}, we present the details of the Spearman correlation result.

As an additional test, we also examined the relationship between total GAB and the satellite fraction of each SAM mock. Interestingly, models exhibiting stronger GAB tend to have higher satellite fractions, although dependent on the selection and number density of the sample. This correlation suggests that satellite populations play an important role in shaping the strength of galaxy assembly bias. In particular, variations in baryonic prescriptions through our SAM can alter the formation and evolution of the satellites, leading to changes in the satellite population and consequently modifying the measured GAB. However, this positive correlation between satellite fraction and GAB does not necessarily imply that satellites are the sole origin of the assembly bias signal, but rather indicates that satellite occupation provides an important channel through which baryonic processes can modulate galaxy–halo correlations.

\section{Discussions and Conclusions}
\label{sec: Conclusions}
In this study, we investigate the drivers of GAB, from its manifestation in OV and its dependence on secondary halo properties, to its underlying connection with the baryonic physics of galaxy formation, using the Galacticus SAM within the UNIT simulation. With hundreds of galaxy mocks that can sample the high-dimensional parameter space, we are able to explore the effect on the GAB signals in a detailed manner.

To quantify the GAB signal, we perform a traditional shuffling method, including both single shuffle and double shuffle. The \textit{single shuffle} is performed by randomizing galaxies among halos of fixed mass, whereas the \textit{double shuffle} further fixes a chosen secondary property, thereby isolating its specific contribution. The GAB signal is then quantified by comparing the two-point correlation functions \(\xi\) of galaxy samples before and after shuffling: we (i) use the single-shuffled sample as the GAB-free reference; (ii) the ratio \(\xi_{\text{original}}/\xi_{\text{single shuffle}}\) yields the total GAB; and (iii) the ratio \(\xi_{\text{double shuffle}}/\xi_{\text{single shuffle}}\) yields the GAB attributable to a particular secondary property.

We first examine the OV and GAB, isolate the contributions of concentration and environmental density, and finally uncover the correlating baryonic processes using a Random Forest machine-learning algorithm.

Our main findings are as follows:
\begin{enumerate}
    \item While environmental density can drive stronger GAB, it does not necessarily drive stronger OV than halo concentration. 


    \item The dependence of GAB on galaxy number density exhibits three distinct types across our SAM parameter space. Type I shows a decreasing clustering ratio with increasing number density for both stellar-mass- and SFR-selected samples. Type II exhibits the opposite trend, with the clustering ratio increasing with number density. Type III shows weak number-density dependence, with clustering ratios remaining broadly stable across the four number-density thresholds. This diversity demonstrates that the impact of secondary properties on GAB is sensitive to the baryonic physics.

    \item The baryonic processes responsible for GAB depend on galaxy number density and the selection criteria. For stellar-mass-selected galaxies, gas cooling and stellar feedback within the disk are consistently the dominant processes, with their relative importance largely unchanged as the number density increases. In contrast, for SFR-selected galaxies, the dominant process shifts from star formation to gas cooling as the number density increases. These rankings are robust against alternative metrics (Spearman, Pearson, SHAP), establishing a direct link between baryonic physics and the GAB.
\end{enumerate}

Our study used the ratio of correlation functions to measure GAB, which was conducted in real space. A natural question is whether peculiar velocities alter this conclusion in redshift space. \citet{Padilla2019} addressed this issue by extending the shuffling technique to velocity space in a different SAM (L-GALAXIES). They found that, for both stellar-mass- and SFR-selected samples, the behaviour of the redshift-space monopole and quadrupole is qualitatively consistent with the real-space correlation function results for GAB and OV reported in \citet{Contreras2019}. Moreover, the structure growth parameter \(f\) inferred from redshift-space distortions remains unbiased by GAB. This suggests that our real-space findings for the baryonic effect are likely to remain qualitatively valid in redshift space, although a dedicated redshift-space analysis with Galacticus would be needed to confirm this.

In this work, we use RF permutation importance to evaluate the contribution of individual baryonic processes to GAB. Another commonly used sensitivity analysis (SA) approach is the traditional one-at-a-time (OAT) method, which evaluates the influence of a single parameter by varying only that parameter while holding all others fixed \citep{Morris1991}. However, the OAT method fails to capture interactions between parameters and has been noted for its limited applicability to high-dimensional, non-linear systems by \citet{Oleśkiewicz2019}, which demonstrates that the variance-based SA method can serve as an alternative to OAT. Here, we analyse SAM mocks generated with Latin hypercube sampling using Random Forest, which naturally models complex interactions among parameters. In the process of computing permutation importance, we have effectively trained a Random Forest regression model that learns the mapping from the 16 baryonic parameters to the GAB amplitude. A natural extension is therefore to directly use this already trained model as an emulator to rapidly predict the GAB strength for arbitrary new parameter combinations, enabling efficient exploration of the high-dimensional baryonic parameter space. The feasibility of such machine-learning-based modeling has been demonstrated, for example, by \citet{Xu2021b}, who used Random Forest to predict galaxy occupation numbers from dark matter halo properties and successfully recovered the GAB signal.

In our analysis of the relative importance of baryonic physics on GAB, the scale of \(10\,h^{-1}\mathrm{Mpc}\) was chosen as a proxy for the strength of GAB, following the recommendation of \cite{Wang2025}, which demonstrates that the fractional change in the clustering signal induced by assembly bias peaks at this scale. It is important to note that they focus only on assembly bias caused by the environment (\(\delta_{10}\)). Our results in Figs.~\ref{fig:ratio}, \ref{fig:ratio_density_Mstar} and \ref{fig:ratio_density_SFR} show that for the environment-induced GAB measured with \(\delta_{10}\), we indeed observe a peak in the clustering ratio at \(\sim 10\,h^{-1}\mathrm{Mpc}\), consistent with \cite{Wang2025}. However, when using a smaller smoothing scale of \(5\,h^{-1}\mathrm{Mpc}\) (\(\delta_{5}\)), the peak shifts to smaller scales closer to \(5\,h^{-1}\mathrm{Mpc}\). This scale-dependent behavior suggests that the characteristic scale of environmental GAB is intrinsically linked to the scale at which the environment is defined. Meanwhile, the concentration-induced GAB and total GAB exhibit peaks at even smaller scales (\(\sim 1\,h^{-1}\mathrm{Mpc}\)), further emphasizing that GAB for different secondary properties manifests at distinct scales. This connection is expected because \(\delta_R\) directly traces the density field on the smoothing scale \(R\), whereas concentration is an intrinsic halo property. But the overall results seem to show that the choice of \(\sim 10\,h^{-1}\mathrm{Mpc}\) is able to provide stable results as an evaluation of the GAB signals.

A related phenomenon that has been discussed in the context of GAB is galactic conformity. It refers to the correlation in star formation activity between central and neighboring galaxies. It was first observed within a dark matter halo (one-halo conformity; \citealt{Weinmann2006}) and later found to extend to scales of several Mpc (two-halo conformity; \citealt{Kauffmann2013}). \citet{Hearin2015} argued that this large-scale two-halo conformity is a direct manifestation of central GAB, and further proposed that it is related to large-scale tidal fields, with the signal weakening at higher redshifts \citep{Hearin2016a}. The observational results of \citet{Berti2017} are consistent with this prediction, supporting the interpretation that two-halo conformity reflects assembly bias. However, alternative explanations have also been proposed, suggesting that conformity may arise from mechanisms other than assembly bias \citep{Paranjape2015,Tinker2018,Zu2018}. On the other hand, using the semianalytic galaxies (SAG) applied to the MDPL2 simulation, \citet{Lacerna2025} showed that the amplitudes of conformity and assembly bias are highly correlated, and both originate from environmental effects around massive dark matter halos. Our results show that GAB signals in the Galacticus SAM are significant on large scales and are primarily driven by environmental density. A feasible test of the connection between GAB and conformity would be to compare the conformity signal before and after shuffling (especially environment-based double shuffling); we defer such investigations to a future work.

Our findings reveal that both the GAB signal and its underlying baryonic drivers differ for stellar mass- and SFR-selected galaxies. These results can provide insights for future parameterizations of GAB in empirical models, particularly regarding the distinction between luminous red galaxy (LRG), emission-line galaxy (ELG) samples and others. 

\begin{acknowledgments}
We thank the anonymous reviewers for their insightful and helpful comments that have significantly improved the robustness of this paper. ZX, JH and ZZ are supported by NSFC (12373003), the National Key R\&D Program of China (2023YFA1605600), and acknowledge the generous sponsorship from Yangyang Development Fund. This work is also supported in part by the China Manned Space Program with grant no. CMS-CSST-2025-A04 and the Office of Science and Technology, Shanghai Municipal Government (grant Nos. 24DX1400100, ZJ2023-ZD-001). The computations in this paper were run on the Gravity Supercomputer at the Department of Astronomy, Shanghai Jiao Tong University.
\end{acknowledgments}

\appendix

\section{HOD Resolution Test} \label{sec:appendix_hod_resolution}

\begin{figure*}
    \centering
    \includegraphics[width=1.0\textwidth]{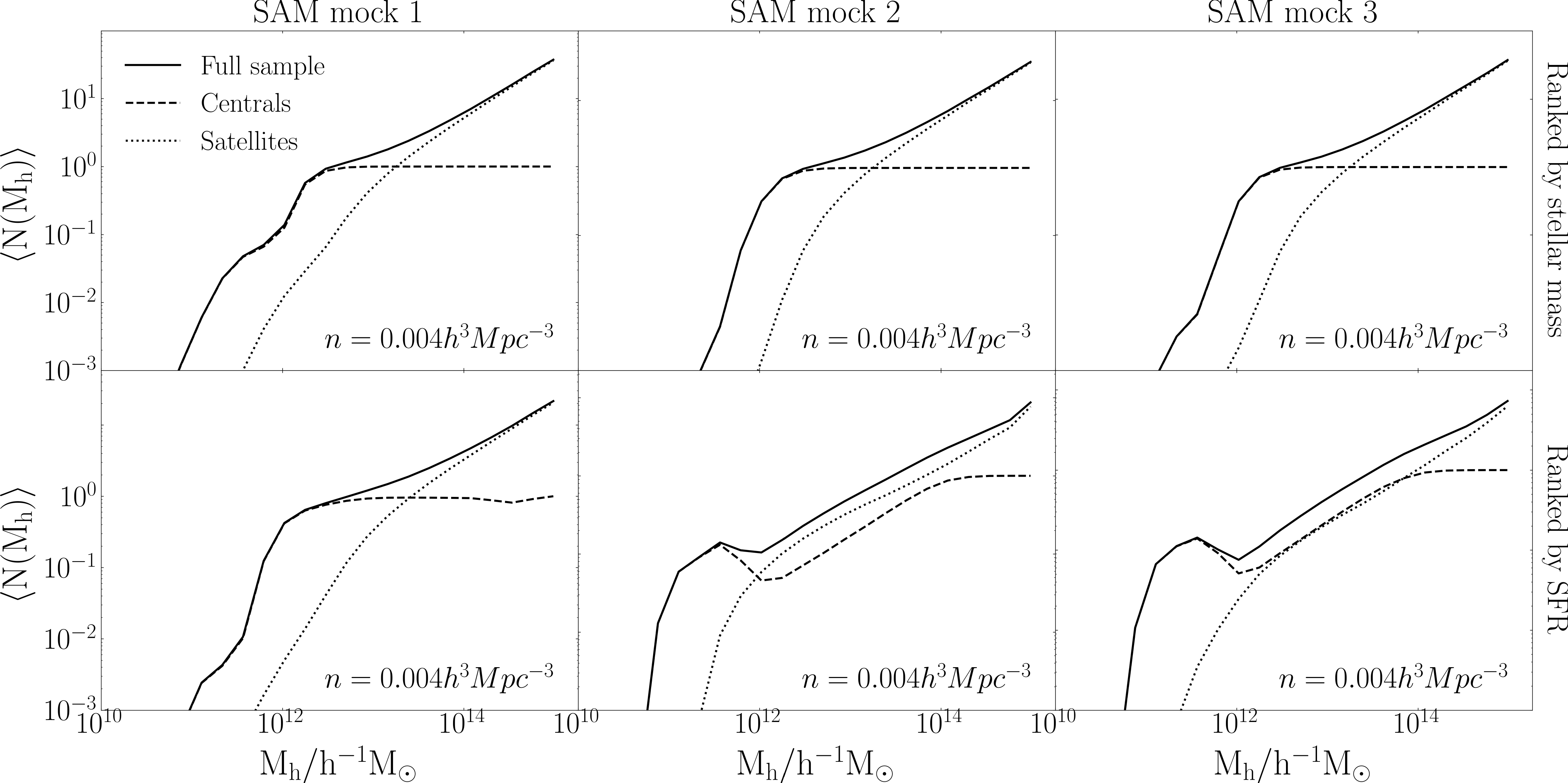} 
    \caption{Halo occupation distributions (HODs) for three SAM mocks that pass our HOD resolution test. These three mocks correspond to SAM mocks 1, 2, and 3 shown in Figs.~\ref{fig:ratio}, \ref{fig:ratio_density_Mstar}, and \ref{fig:ratio_density_SFR}. The columns show the three mocks from left to right. The top row shows galaxies selected by stellar mass, while the bottom row shows galaxies selected by SFR. In each panel, the HOD is decomposed into central and satellite contributions.}
    \label{fig:HOD}
\end{figure*}

\begin{figure*}
    \centering
    \includegraphics[width=1.0\textwidth]{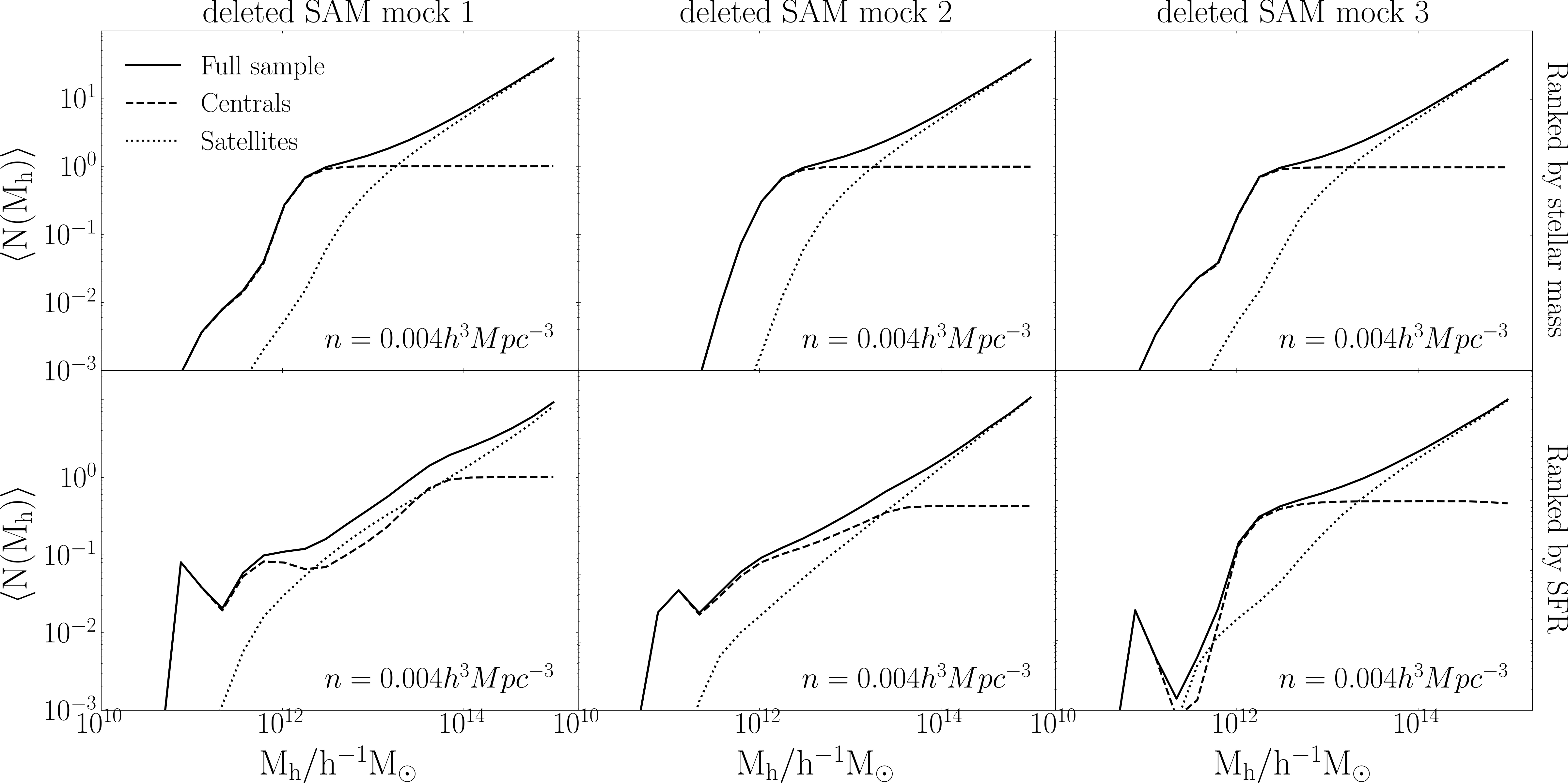} 
    \caption{Same as Fig.~\ref{fig:HOD}, but for three representative SAM mocks removed by the HOD resolution test. The columns show deleted SAM mocks 1, 2, and 3 from left to right.}
    \label{fig:HOD_deleted}
\end{figure*}

We inspect the HODs of all the SAM mocks used in this work in order to identify samples affected by resolution-driven features at the low-mass end. The original catalogue set contains 262 Galacticus mocks. In the SFR-selected samples, an unphysical bump at \(M_h < 10^{11}\,h^{-1}M_\odot\) can appear when the target galaxy number density is too high, forcing an artificially large occupation in the least massive halos below the reliable resolution limit. We therefore identify mocks with this low-mass HOD bump as resolution-affected and remove them from the subsequent analysis. This screening leaves 133 mocks. To avoid comparing Random Forest results trained on different model subsets, we use this same set of 133 HOD-selected mocks for all the selection criteria. Fig.~\ref{fig:HOD} shows the HODs of three retained mocks, corresponding to the SAM mocks 1, 2, and 3 used in Figs.~\ref{fig:ratio}, \ref{fig:ratio_density_Mstar}, and \ref{fig:ratio_density_SFR}. For comparison, Fig.~\ref{fig:HOD_deleted} shows three representative mocks removed by this resolution test.

\section{Complete Random Forest Feature Importance} \label{sec:appendix_rf} 

\begin{figure*}
    \centering
    \includegraphics[width=1.0\textwidth]{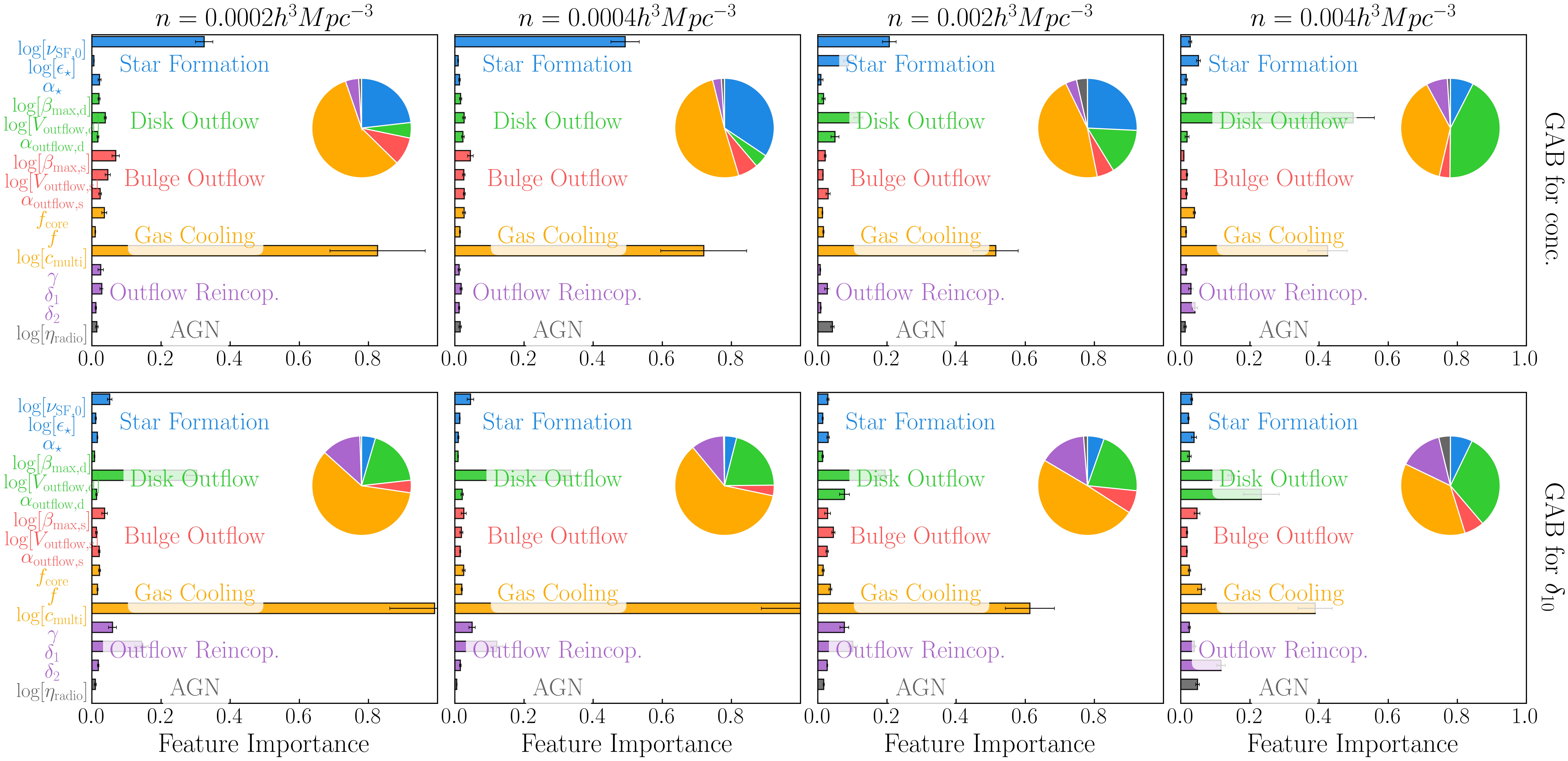} 
    \caption{The relative importance of various baryonic processes on GAB for stellar-mass-selected galaxy samples, using the same plotting conventions as Fig.~\ref{fig:feature_importance}. From top to bottom, the rows show the feature importance for concentration-related GAB and \(\delta_{10}\)-related GAB. From left to right, the columns correspond to increasing galaxy number density.}
    \label{fig:feature_importance_Mstar}
\end{figure*}

\begin{figure*}
    \centering
    \includegraphics[width=1.0\textwidth]{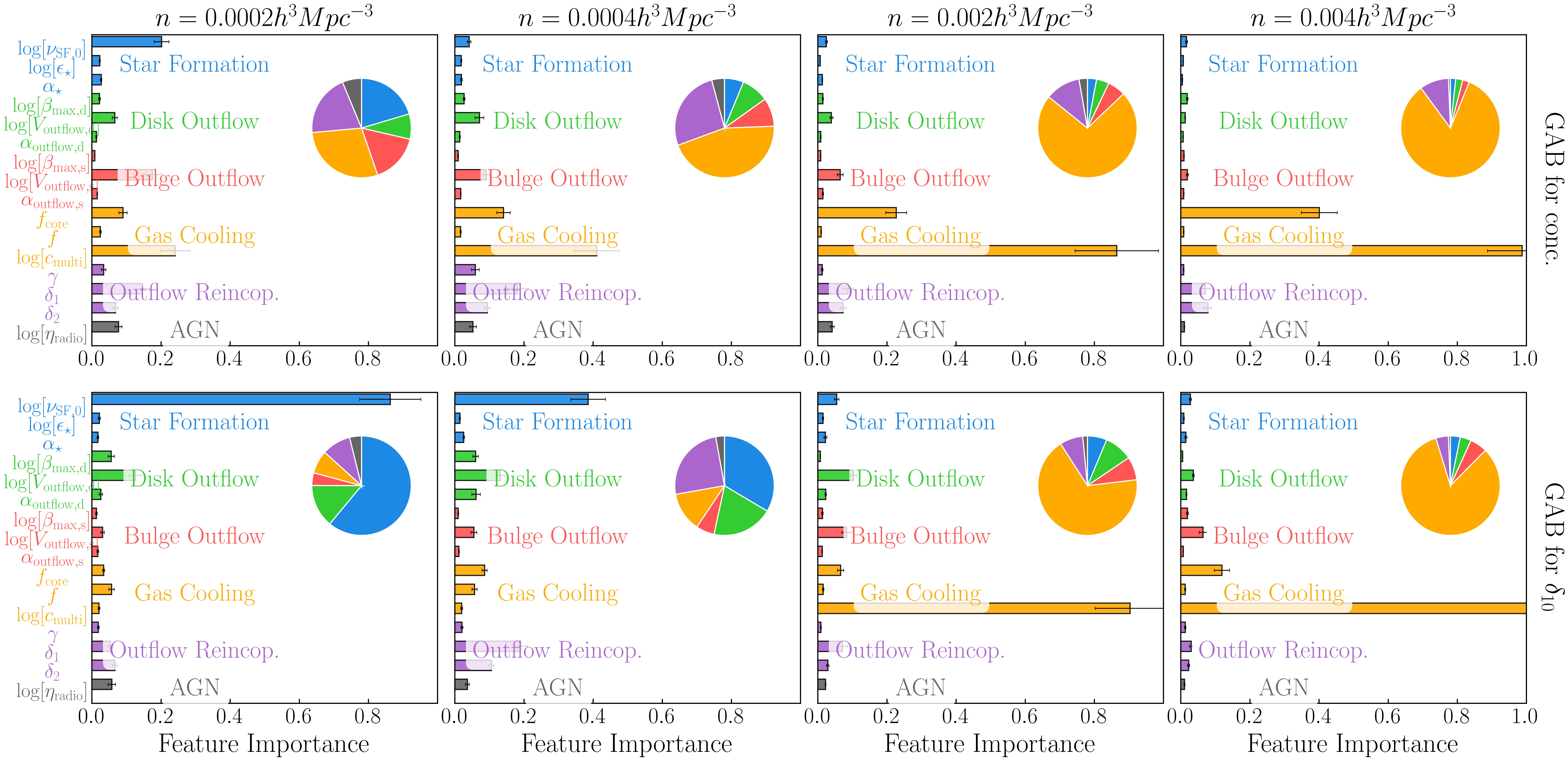} 
    \caption{Same as Fig.~\ref{fig:feature_importance_Mstar}, but for SFR-selected galaxy samples.}
    \label{fig:feature_importance_SFR}
\end{figure*}

Figs.~\ref{fig:feature_importance_Mstar} and \ref{fig:feature_importance_SFR} show the additional RF feature-importance results, including GAB associated with concentration and \(\delta_{10}\)\footnote{The result with $\delta_{5}$ is very similar to $\delta_{10}$}. The feature-importance patterns for the environment-related GAB signals, \(\delta_{10}\) and \(\delta_{5}\), are broadly consistent with those for the total GAB signal. The concentration-related GAB shows some differences at low number density. For stellar-mass-selected samples, star formation parameters become relatively more prominent in the concentration-related measurement, while for SFR-selected samples, gas cooling remains more important. Nevertheless, the main conclusions are unchanged: for stellar-mass-selected samples, gas cooling is the most important process across the number densities considered here, while for SFR-selected samples, the importance of gas cooling increases with number density and eventually becomes dominant.

\section{Spearman Correlation Analysis} 
\label{sec:appendix_Spearman}

\begin{figure*}
    \centering
    \includegraphics[width=1.0\textwidth]{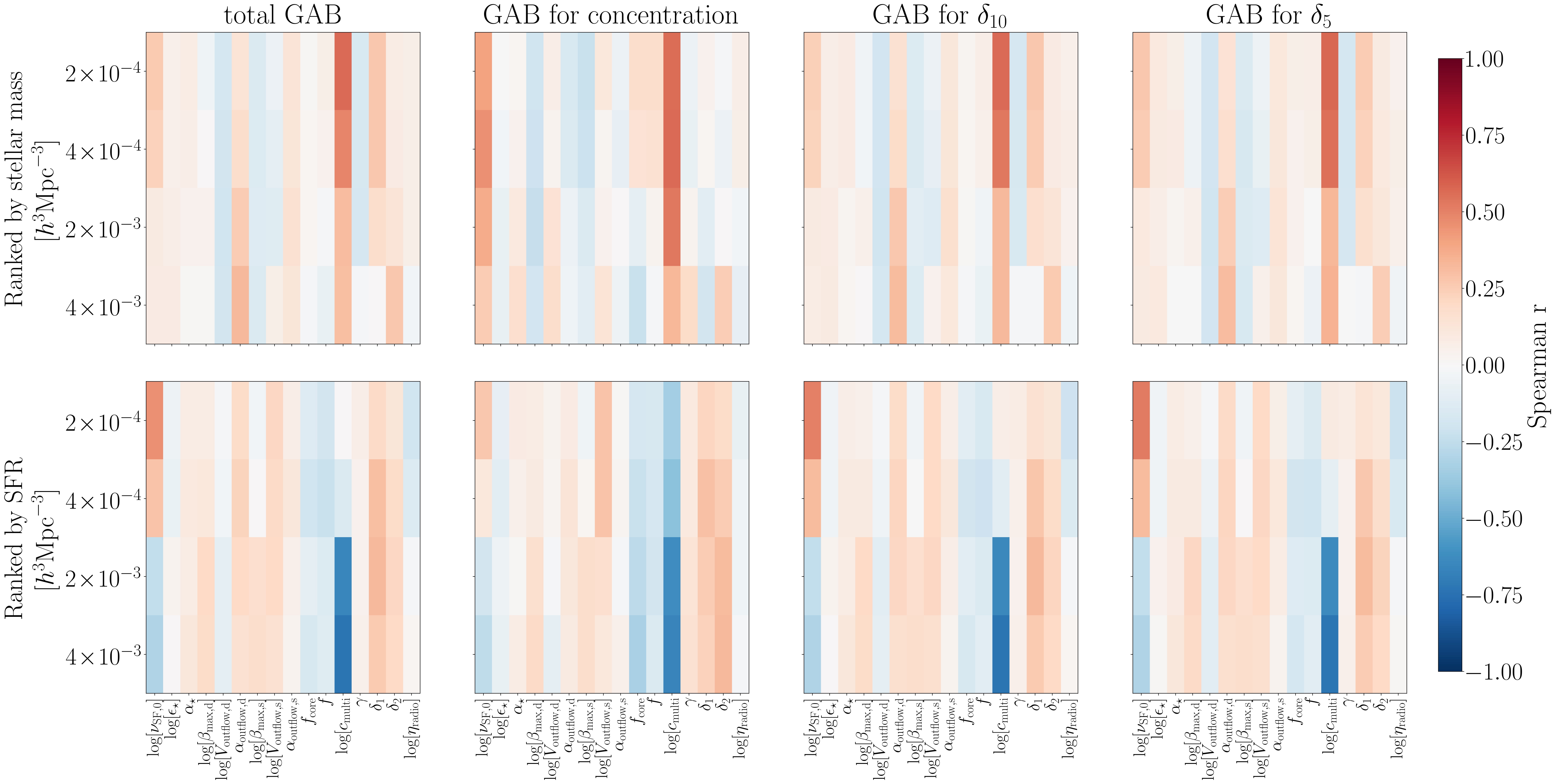}
    \caption{Spearman correlation coefficients between baryonic processes and GAB for stellar-mass-selected (top) and SFR-selected (bottom) samples. The y-axis shows four galaxy number densities, and the x-axis lists the SAM parameters. Red (positive) and blue (negative) colors indicate the direction and strength of correlation.}
    \label{fig:spearman}
\end{figure*}

Fig. \ref{fig:spearman} shows the Spearman correlation coefficients between each SAM parameter and the GAB signals. Each subplot shows the correlation strength across four galaxy number densities and 16 SAM parameters. The color bar indicates the Spearman \(r\) value, where red represents positive correlation, and blue represents negative correlation. For stellar-mass-selected samples, consistent with the Random Forest analysis, \(c_{\mathrm{multi}}\) exhibits the strongest positive correlation with GAB across all number densities compared to other parameters. For SFR-selected samples, the correlation patterns also align with the Random Forest analysis, reflecting the transition from \(\log[\nu_{\rm SF,0}]\) dominance to \(c_{\mathrm{multi}}\) dominance as number density increases. We also perform similar tests using methods including Pearson correlation and SHAP analysis, and they can yield similar results.

\bibliographystyle{aasjournalv7}
\bibliography{reference}

\end{document}